# Do Species Evolve Through Mutations Guided by Non-Coding RNAs?

**Reza Rahmanzadeh[1][#]**

[1]Support Center for Advanced Neuroimaging (SCAN), University Institute of Diagnostic and Interventional Neuroradiology, Inselspital, Bern University Hospital, University of Bern, Bern, Switzerland.

# Corresponding author – Reza.Rahmanzadeh@insel.ch

**Abstract**

The current theory of evolution is almost the one Darwin and Wallace proposed two centuries ago and the following discoveries e.g., Mendelian genetics and neutral mutation theory have not made significant modifications. The current evolution theory relies mostly on heritable variations within species population, natural selection and genetic drift.

The inability of the current theory to explain and predict biological observations, especially the emergence of evolutionary novelties, highlights the need to incorporate recent evolutionary, developmental and genetics findings in order to achieve a more comprehensive explanation of species evolution.

The present paper provides significant body of evidence to substantiate a new theory to account for species evolution. The main axes of the proposed theory include:

First, mutations leading to genetic novelties in a given species during evolution should be guided by the environment surrounding that species. Second, environment and germline are connected with each other via soma-germline messengers e.g., non-coding RNAs (ncRNAs). Third, based on the information that germline continuously receives in terms of epigenetic messengers three stages of heritable changes may occur in germline genome to produce more adaptable offspring: i, Epigenetic modifications, ii, Genetic mutations in the sequence of pre-existing genes in order to improve their potency, and iii, The production of new genes with distinct gene-coding regions, which may have their own regulatory regions.

Almost two hundred years after Darwin and Wallace, the dominant theory accounting for the origin of species is basically almost unchanged and relies on existing heritable variations in species population, natural selection and genetic drift (1).

This paper tries to have an independent look at evolutionary processes and to incorporate recent findings, especially genetic and epigenetic ones, to reformulate principals underlying the evolution of species. I will start with an extreme example and then try to write main axes of the proposed theory into distinct sections.

## I) First Time for a Species to Experience a Stress

Consider a plant species million years ago as our example. And consider the time at which for the first time in the history of the world a given stress will be experienced by a species, in other words no species by that time experienced that given stress.

For example, assume that the soil salt concentration in the habitat of a plant species is 3 g/l and it's changing to 100 g/l over a very short period on the evolutionary time scale. As it's the first time that this species will be exposed to new concentrations of salt, therefore it's not far away from reality to consider that the salt concentration that this plant species could tolerate at its best (with recruiting all its compensatory potency) to be something lower than 100 g/l, let's consider 15 g/l. Now what will happen following exposure to this stress? Based on the current theory of evolution, there should be genetic material in some species individuals enabling a new function which is to tolerate the 15-100 g/l changes in salt concentration, otherwise the species may be extinct. Hereafter, we refer such new functions in some species individuals as to "phenotypic novelties", which may be selected by natural selection later in the course of evolution and become fixed in the species.

## II) How Probable It Is to Harbor A Specific Phenotypic Novelty at Stress Time?

i, Based on the current concept, genetic novelties underlying phenotypic novelties on which natural selection may act are products of random mutations. It means that this random process may give rise to any phenotypic novelties without any priorities. In theory, the quantity of such phenotypic novelties for a given species is limitless, from small to big changes in the quantity and quality of preexisting phenotypes to new non-existing ones which again could theoretically be limitless. As a result, the likelihood for this species to harbor this specific phenotypic novelty to survive the 15-100 g/l change in soil salt concentration in some individuals at the time of stress is very close to zero.

ii, One might argue that considerably long time like what we have in the evolution of species could increase the likelihood for necessary phenotypic novelties to evolve. Even having much more time may not be a solution. Given that this stress was not experienced till now by assumption, therefore there was no selective advantage for the species individuals to carry this genetic novelty (in this case making them able to tolerate 15-100 g/l soil salt change). As a result, long time not only help species to develop new genetic and phenotypic novelties, but also to lose them due to the lack of current advantages.

iii, Moreover, the likelihood for species individuals to develop and collect all necessary genetic materials bringing about a phenotypic novelty could not be considerable. The reason comes from the fact that to have a phenotype at least a group of genes with well-tailored regulatory regions, e.g. promoter and enhancer, and with other regulatory genes (In this paper we will refer such a collection of genetic materials as to "genetic novelty"; **Box 1. Regulatory Network for Gene Expression**) have to be present in the genome altogether. Recent studies (2-4) show that even an abnormality in one regulatory sequence could cause devastating disorders that may reduce survival and fitness. As a result, gathering of all these genetic materials to bring about a phenotypic novelty that natural selection may act on is very unlikely given that lacking one of these materials not only prevent the emergence of the phenotype, rather could reduce fitness by causing disorders.

Taken together, it seems that there should be a priority for some phenotypic novelties, otherwise due to countless quantity of potential phenotypic novelties the extinction rate of species upon encountering stresses for the first time will be extremely high. The section below provides more supports for that.

### III) Chemical World vs. Biological World: Dependent?

Consider the world before the emergence of the first biological entities. In this paper, we will call the biological entities and organisms that undergo biological evolution as "biological world" and the reset of the world, that was also present prior to the emergence of the biological world, as "chemical world". All the chemical world, and of course the biological one, is composed of chemical elements that combine together to form chemical compounds with distinct physical and chemical properties. In this paper we will refer chemical compounds with unique properties as to chemical novelties. The biological world anyway has to be built based on these chemical novelties as there is no other resources beside them to be consumed by the biological world. For example, water (a combination of two hydrogen atoms and one oxygen one), salt (a combination of one sodium and one chloride atoms) and oxygen (a combination of two oxygen atoms) all are indispensable chemical novelties for biological organisms. Therefore, this biological world has to be built in a way to be dependent only on these existing chemical novelties and nothing else.

Now consider the current evolution theory that all genetic novelties will happen randomly. As the mutations and genetic novelties are not confined to be produced in a way to become compatible with the existing chemical novelties (the environment surrounding a given species), therefore they will make genetic novelties that underlie phenotypic novelties usable in one of the limitless imaginable chemical worlds, for example imagine a world in which xenon is abundant enough to play the same role as oxygen in our world for biological species. And as a result, the likelihood for the plant species of our example in the previous section to harbor phenotypic novelty for 15-100 g/l salt concentration is almost zero as theoretically limitless quantity of phenotypic novelties, i.e. countless number of possible chemical world multiplied by countless number of possible phenotypic novelties in a given chemical world, could have the chance to be present in some individuals at the time of stress.

The sections **I-III** suggest that genetic novelties should be created in a way to be applicable for a species living in a special chemical world and, in this regard, mutations leading to genetic novelties should be guided by the environment surrounding a given species. In the sections below I will try to take advantage of novel epigenetic discoveries to show how environmental experiences and stresses might underlie gene modifications in species.

---

**Box 1. Regulatory Network for Genes Expression**

Consider that a new gene (both regulatory regions, e.g. promoter and enhancer, and coding region) is formed during the evolution. This gene could not properly function unless all necessary regulatory genes (See Glossary below) containing instructions of "how to use" are present in the genome. These instructions determine when, where and how much gene expression needs to happen (Sections a and b below). Furthermore, this gene should be correctly wired with other genes in order to ensure their cooperation to produce a function (Sections c and d below).

The necessary regulations of a given gene include:

a) The baseline amount and fluctuation range of gene expression in physiological conditions. During the development, some regulatory factors will determine the tissue-specificity of and the baseline amount of gene expression. Such regulatory factors should have sequence-specificity and the ability to recruit chromatin modifiers to implement a stable epigenetic mark (transcriptional regulation). Transcription factors and long non-coding RNAs (lncRNAs) have been proposed to bring about this function (5, 6). In addition to regulating the baseline level of gene expression, there should be other regulatory factors causing normal fluctuations in gene expression due to dynamicity of physiology. The best candidate for this function are micro RNAs (miRNAs) and small interfering RNAs (siRNAs) that could regulate gene expression via post-transcriptional mRNA cleavage and translational silencing (7, 8).

b) In non-physiological conditions as environmental stresses, there is usually a need to change the expression of some genes for short- or long-term (9, 10). In such conditions, gene expression may exceed the physiological ranges. Almost all ncRNAs, especially miRNA and siRNA have been shown to rapidly change following stresses (11, 12). ncRNAs via post-transcriptional and transcriptional regulations could mediate short- and long-term changes in gene expression (13). Regulatory factors that mediate long-standing transcriptional regulation during non-physiological conditions should have some features: accumulation in tissues that are targeted by stress, sequence-specificity in order to find corresponding genes and the ability to recruit chromatin modifiers. For example, in plants siRNAs are such regulatory factors (14, 15).

c) Functions are usually produced by a group of genes working together. Therefore, the expression of a given gene should be tuned concertedly with other genes (i.e., horizontal wiring). It means that for instance a group of co-working genes might harbor sequence similarities, e.g. transcription factors (TFs) binding sites, in their regulatory regions making them able to be concertedly expressed by a group of TFs. In addition, similarities in their regulatory sequences to synchronize the transcriptional and post-transcriptional regulation of co-working genes. And a given gene is usually involved in a wide range of functions in each of which it works with a distinct group of genes. It means that a given gene should harbor different sets of sequence similarities, e.g., TF binding sites, in its regulatory regions allowing it to be horizontally wired with distinct sets of genes.

d) The expression of a given gene may be triggered and influenced by other genes and, in turn, a given gene may influence the expression of others (i.e., vertical wiring).

Therefore, one could conclude that a couple of regulatory factors (e.g., ncRNAs) and regulatory regions (e.g., promoter and enhancer) will shape the regulatory network controlling gene expression. For ncRNAs sequence-complementarity with target genes and for regulatory regions sequence similarities (e.g., TFs motifs) with the regulatory regions of other co-working genes make concerted regulation of genes possible. All these elements providing sequence similarities and complementarities that are embedded within regulatory regions of genes or ncRNAs targeting the genes, which could be shared among co-working genes in order to synchronize their activation and expression, will hereinafter be called "regulatory elements" - to recall chemical elements in chemistry. The author hypothesizes that regulatory regions and ncRNAs contain regulatory elements the same with chemical compounds that are made out of chemical elements. Such elements-based recognition may underlie complex regulatory network in the biological world.

Furthermore, both regulatory regions and regulatory ncRNAs may have similar sequence origins (16-23). Especially, it is shown that transposable elements (TEs) harbor several TF binding sites and most of TF binding sites reside in several TEs (24-30). In addition, it is shown that TEs are extensively present in ncRNA genes (e.g., miRNA and lncRNA) and some other ncRNAs e.g., siRNA originate mainly from TEs (11). For example, tandem repeat element present in the promoter of a sodium transporter will be targeted by small ncRNAs mediating methylation and gene suppression. Therefore, repetitive elements may contain myriads of regulatory elements present in regulatory regions and ncRNAs.

**Glossary**

**Regulatory regions:** genetic regions upstream, downstream or within the genes involved in the regulation of gene expression. Promoter and enhancer are the best known of them and harbor TF binding sites.

**Regulatory sequences:** All ncRNAs involved in transcriptional and post-transcriptional regulation of gene expression e.g. miRNA, siRNA, lncRNA. Some of them as lncRNAs and miRNA have their own genes.

**Regulatory factors:** In addition to ncRNAs, this includes also regulatory proteins as TFs.

**Regulatory genes:** genes encoding regulatory factors or their precursors.

## IV) How Organisms React to Environmental Conditions?

Organisms respond to environmental conditions e.g., stresses via altering and regulating the expression of couple of genes in targeted organs or tissues (**Box 1. Regulatory Network for Gene Expression**). Various kinds of regulatory factors are involved in the post-stress gene expression regulation and ncRNAs have been shown to play an axial role (11, 12, 22, 31). miRNAs are involved in post-transcriptional regulation and directly target the mRNAs of genes to cleave or to suppress translation. Organisms usually change the expression of certain miRNAs in the early phases of stresses (12, 15). For example, upon osmotic stress in zebrafish, microRNA miR-8 represses the expression of a transporter to regulate Na/H exchange across membrane (32). siRNAs more than post-transcriptional regulation may also alter the transcriptional gene expression via recruiting chromatin modifiers to the location of genes. siRNAs through mediating DNA and histone modifications may alter gene transcription for a longer period (33). For example, in Arabidopsis thaliana grown in the environment with low relative humidity, siRNAs associated with FAMA and SPEECHLESS gene locus, which are two genes with axial function in stomatal development, will be increased and lead to siRNA-dependent DNA methylation of the genes (34). As a result, siRNA reduces the transcription of these genes and stomatal frequency, thereby plants grown in low humidity could keep water during this stress. The expression of lncRNAs are also elevated during stress and may alter gene expression transcriptionally through making an RNA scaffold and the recruitment of chromatin modifiers (35). Evidence shows lncRNA concentration changes following stress (31). Although there is a lot of evidence showing prominent involvement of ncRNAs during stresses, they are the main gene expression regulators in physiological conditions as well (Box 1. Regulatory Network for Gene Expression).

Moreover, an increasing body of studies shows that parental environmental experiences could be transferred to offsprings through "epigenetic inheritance" (36-41). This means that changes in gene expression, e.g. transcriptionally or post-transcriptionally, in the somatic tissues of the parent experiencing environmental stress could somehow be transferred to the germline and next generations (42- 45). Although the messengers (**Box 2. Messengers**) between soma and germline are not yet fully recognized, the existence of the considerable amount of mobile and vesicle-embedded ncRNAs in body fluids suggests those as the best candidates.

Taken together, recent findings show that organisms react to environmental changes and stresses through regulating gene expression, a process that is mediated by regulatory factors especially ncRNAs. And, ncRNAs have the ability to go to germline and to regulate the expression of corresponding genes in next generations. This soma-germline route will make a bridge via which environment may influence germline genome.

---

**Box 2. Messengers**

When a regulatory factor could be called soma-germline messenger?
i, Increased concentration in target somatic tissues following environmental alterations (e.g., stresses)
ii, An axial function in the regulation of gene expression in somatic tissues following environmental alterations
iii, Evidence for mobility, freely or in the context of extracellular vesicles, from soma to germline.
iv, Evidence for potential involvement in trans-generational epigenetic inheritance.

Recent studies suggest ncRNAs (e.g. miRNA, siRNA, piRNA, tsRNA and lncRNA) and in some species some proteins as excellent candidates for soma-gremline messengers (46, 47). However, there may be differences among different species, for example tsRNAs may be more involved in mammals and siRNAs more in plants. In this paper we will consider these potential candidates as soma-germline messengers.

## V) The Theory of Species Evolution Through Guided Mutations

The sections I-III show that environment should somehow affect genetic mutations and novelties, otherwise species evolution could not effectively continue. The section IV links environment to germline via messengers e.g., ncRNAs.

Any given organism experiences continuously the environment including a stable environment or environmental instabilities (e.g., stresses). From the organismal viewpoint, this experience will be encoded in terms of regulatory factors controlling and altering the expression of genes. Although regulatory factors are not yet fully understood, ncRNAs are emerging as the prominent ones (section IV).

Changes in the concentration of regulatory factors in somatic tissues will be sent to germline via messengers. It means that always in organism life, germline is able to see the full picture of environment through the glass of messengers originating in and departing from somatic cells.

This theory proposes that based on the information that germline continuously receives, three stages of heritable changes (which will be totally called "Heritable Gene Changes in Germline"; HGCG) may occur in germline genome to produce more adaptable offsprings:

HGCG-i, Epigenetic modifications; it means that following exposure to stress, alteration in the baseline expression of some genes is needed and this change is yet sufficient to respond to the environmental stress.

If changes in the baseline gene expression could not meet organism need to respond to the environment, then genetic mutations will occur through ii & iii.

HGCG-ii, Genetic mutations in the sequence of pre-existing genes in order to improve their potency, e.g., gain-of-function point mutations (this item is the only one in which random mutation and genetic drift - as the current theory suggests - may play a significant role and, therefore, to avoid unnecessary discussion will not be covered in this paper).

HGCG-iii, The production of new genes with distinct gene-coding regions. Those may have their own regulatory regions and regulatory genes.

How messengers could trigger three-stage heritable changes in the germline?
**Box 1. Regulatory Network for Gene Expression** shows and hypothesizes that ncRNAs and regulatory regions related to genes are composed of some regulatory elements. **Box 2.**

**Messengers** shows that soma-germline messengers are mainly ncRNAs which are produced in somatic cells. Messenger ncRNAs traveling to germline are produced in somatic cells where they are able to change the expression of a couple of genes in response to environment. These ncRNAs harbor regulatory elements and could recognize their corresponding sequences in germline and therefore could manage the sequence-specificity of the HGCG.

For epigenetic modifications (HGCG-i), the presence of messengers itself could lead to post-transcriptional changes in a number of next generations. For example it has been shown that starvation stress in Caenorhabditis elegans will change the expression of small RNAs targeting the genes with roles in nutrition, an effect that lasts over three generations due to transgenerational inheritance of small RNAs (48). However for longer period of epigenetic modifications, ncRNAs should recruit chromatin modifiers to implement transcriptional changes. However, these changes in gene expression should be implemented in a way to alter the expression of target genes only in target cells or tissues of offsprings. For example it has been shown that odor fear conditioning could be transgenerationally inherited over three generations and will increase the quantity of olfactory sensory neurons expressing receptors for that specific odor in the main olfactory epithelium (49). The theory proposes that the collection of ncRNAs originating from the same somatic location altogether will manage the target-specificity of epigenetic modifications.

To show how messengers could trigger the production of genetic novelties (HGCG-iii), we will consider a plant species (of section I), given that in plants the messengers are better investigated. However, based on the proposed theory the main principles underlying HGCG-iii will be quite similar in all species.

For making a couple of genes, we should consider the situation that changes in the expression level of existing relevant genes are not sufficient for the species to survive the new stress and, therefore, new genetic materials (new genes or mutations) are needed to be produced. Therefore, messengers in germline represent ncRNAs regulating the expression of genes that would-be-produced genes should be wired with since those all should serve similar (not fully equal) functions.

Regulatory elements within the regulatory regions and ncRNAs shape the regulatory network in organisms. Therefore, regulatory regions for new genes should be built in a way to contain regulatory elements (some of those are shared with existing genes, which the new gene should be wired with) in order to be incorporated into the regulatory network.

In plants, most messengers are siRNAs and most siRNAs in plants originate from transposable elements (TE). Therefore, these siRNAs (messengers) may be able to alter the expression of TEs and TE- associated genes in germline. As a result some TEs may find this chance to escape their tight repressing controls, which is mediated by vegetative cells, to get expressed in germline. Now consider that these siRNAs themselves and/or of activated TEs (or parts of them) as regulatory elements may be incorporated into the regulatory regions and regulatory genes of new genes. In addition, messengers may be able to activate new TEs and re-wire simultaneously co-working genes (50, 51). As a result, through this theory new genes will automatically be wired with other relevant genes in regulatory network.

Studies attribute new gene formation (coding regions) to a few main mechanisms (52- 54). First, gene duplication that generates new copy of a given gene via DNA-based duplication or through making a new intron-less copy out of gene mRNA by reverse-transcription (55, 56). Second, chimera formation that makes new genes via combining genes, domains or exons of existing genes (57,58). Third, de novo gene formation produces new genes that are not made using existing genes (59- 61). These genes in some species are usually composed of existing non-coding sequences. Another group of genes are ncRNAs and their origins were discussed in Box 1.

Now consider that following an environmental exposure (e.g., increase in soil salt concentration) the expression of siRNA associated with genes involved in relevant pathways (e.g., pathways involved in the regulation of sodium uptake and intracellular salt concentration) will change. This change in siRNAs will be sent to germline. The concentration and profile of siRNAs coming from somatic tissues into germline and also the frequency for a given siRNA to go to germline may altogether guide HGCG towards one of HGCG-i, -ii and -iii.

Now consider a situation that this collection of siRNAs in plant germline signifies that the existing genes even at their most favorable regulatory status could not respond fully to stress and there is a need for mutations in the existing genes (HGCG-ii) or for making new genes (HGCG-iii). In this case, the sequence specificity of ncRNAs orients HGCG towards one of the gene formation mechanisms e.g., gene duplication. siRNAs may be able to trigger duplication (DNA-or mRNA-based) through potential changes in the activation of TEs (which these specific siRNAs originate from) or through the expression of TE-associated genes or corresponding genes themselves (in this case, genes involved in salt pathways) in germline.

Now consider that there is a need for the organism to express one gene in a new location (e.g., another cells or tissue compartment) beside the tissues usually express it. Then some ncRNAs from this new location go to germline to forge the regulatory regions of the new copy of the gene (that is usually a RNA-based copy and could be produced through above mentioned methods) in a way to be expressed in the new location. This explanation could also be generalized for chimeric structures formation and de novo ones that will be produced by the combination of protein-coding genes and non-coding sequences. However, the situation is by far more complicated in de novo gene formation through which protein-coding genes and novelties will be produced from non-coding sequences. The guided-mutation theory hypothesizes that HGCG could make coding regions out of non-coding ones, a process that will be triggered and guided by messengers (see the section chemical world's novelties). Guided-mutation theory considers a prominent role for TEs as one major effector of HGCG, especially in HGCG-iii. Indeed, an increasing body of data supports TE role in evolution. Previous studies (63- 66) show the role of TE in regulatory sequence formation and in all main mechanisms for gene formation.
For HGCG-iii, target specificity will be determined by regulatory genes that will be generated in parallel to a protein- coding genes.

### VI) Core Theory

The theory of "species evolution through guided-mutations" proposes that mutations underlying the emergence of novelties could not be produced through random blind processes as the current evolution theory proposes. As discussed in the sections I-III, environment should guide mutations in a way that genetic novelties more applicable for the species to live in the surrounding environment having more chance to be produced. In more general words, the biological world will not be produced in the vacuum and will be produced in a pre-existing chemical world and will be built in a way to use chemical world novelties as resource.

An individual in a given species experiences continuously its surrounding environment. All cells of the individual are somehow experiencing the environment and responding to it in terms of regulatory factors (especially ncRNA) to regulate the expression of genes. The ncRNAs departing from all cells go to germline and simulate the environment in the germline. As messenger ncRNAs and the regulatory regions of genes are wired together in the regulatory network, the ncRNAs convey alot of information about the changes in gene expression in somatic tissues to germline. Collectively, germline will have a complete picture of environment (or at least of environmental alterations). The ncRNAs have the ability to trigger and manage three stages of heritable gene changes in germline (HGCG); i-epigenetic modifications, ii-mutations to improve function of existing genes and iii-formation of new genes (with distinct coding region). The profile, concentration and frequency of all ncRNAs - originating from various somatic tissues and cells - in germline will determine the outcome of HGCG. Therefore, the author hypothesizes that there is a system in germline that its inputs are messengers (especially ncRNAs) and its output is well-tailored HGCG. This HGCG system tries to change parental germline genome in a way to make offsprings more adaptable to the surrounding environment and therefore will increase the overall survivability and fitness. It's hard to predict how fast HGCG system works in a species individual, however the author thinks that the speed depends on the severity and frequency of a given environmental cues that the species individuals experience, which will be transferred to germline in terms of the concentration, frequency and profile of ncRNAs. And the time takes for the germline to implement HGCG is very short on the evolutionary time scale and may occur in the individual (or the generation) that for the first time experiences the stress. Of note, the guided-mutation theory does not propose that all mutations will occur following experiencing the environmental events, especially HGCG-ii mutations, rather it proposes that mutations more applicable and useful for a given species in a given environment are much more likely to happen and environmental experience encoded as ncRNAs will guide occurance of such applicable epigenetic/genetic alterations.

### VII) More Supports for the Guided-Mutation Theory

i, Convergent novelties show that evolution responds to the same questions (e.g., environmental exposures) with the same solutions (e.g., novelties) (67, 68). These repeated input-output pairs might further notify the presence of a system (as HGCG) to be responsible for emergence of novelties.

Of importance, genes that are produced independently more than one time during evolution show that different gene copies are usually made out of similar already existing genetic pieces, which combine together with similar gene formation mechanisms (see Table 1). This finding might suggest the presence of a couple of regulations in new genes formation.

For example, TRIM5-CypA chimeric gene was independently generated in New World monkey and Old World monkey due to CypA retroposition into the TRIM5 gene and deletion of the original capsid binding domain (B 30.2). The TRIM5-CypA chimeric gene is more potent against HIV1 and other retroviral infection (69). Table 1 summarizes some examples of convergent gene formation with similar processes.

ii, Evidence shows that newly evolved genes are functional and even essential for species. This evidence may further suggest that new genes may originate functional or in a way to get quickly functional (70, 71).

iii, In the previous studies the correlation between gene age and functions is investigated. This correlation shows that genes involved in similar functions might be produced not far away from each other in evolution. It again signifies the importance of a forced mutations in favor of one function at a given time in evolution (72- 76).

**Table 1.** Convergent gene formation in distinct species/lineages with similar mechanisms

| Gene | Species/ Lineage | Mechanism of gene formation | New function |
|---|---|---|---|
| RAG1-RAG2 (77) | Purple sea urchin Green sea urchin | Transib transposable element | Essential subunits of V(D)J recombinase |
| Mammalian CENP-B (Abp1, Cbh1, Cbh2 in fission yeast) (78) | Mammals Fission yeast | Pogo-like transposons | Chromosome segregation, hetrochromatin formation |
| Ntf-2 and ran (79) | Drosophila melanogaster, D. ananassae, and D. grimshawi | X-to-autosome retroposition | Transportation of proteins into and out of the nucleus |
| Adh-derived chimeric genes (jingwei, Adh-Finnegan, Adh-Twain) (80) | Dorsilopha genus (each gene generated independently in multiple species) | Adh retroposition and fusion with 5' end of another genes | Relative to Adh, the substrate range is expanded |
| TRIM5-CypA (69) | Old World Monkey New World Monkey | CypA retroposition into the TRIM5 gene and deletion of the original capsid binding domain. | More potent against HIV1 and other retroviral infectio |